\documentclass{PoS}
\usepackage{lineno}

\title{The H.E.S.S. II GRB Program}

\ShortTitle{The H.E.S.S. II GRB Program}

\author{\speaker{R.D. Parsons} \\
Max-Planck-Institut f{\"u}r Kernphysik, Heidelberg, Germany\\
E-mail: \email{daniel.parsons@mpi-hd.mpg.de}
}
\author{ A. Balzer\\
GRAPPA, Anton Pannekoek Institute for Astronomy,
University of Amsterdam, Amsterdam, The Netherlands\\
E-mail: \email{arnim.balzer@uva.nle}
}
\author{ M. F{\"u}ssling\\
 DESY, Zeuthen, Germany\\
E-mail: \email{matthias.fuessling@desy.de}
}
\author{C. Hoischen\\
Institut f\"ur Physik und Astronomie, Universit\"at Potsdam,
Potsdam, Germany\\
E-mail: \email{clemens.hoischen@desy.de}
}
\author{{M. Holler}\\
Laboratoire Leprince-Ringuet, Palaiseau, Ecole Polytechnique,
CNRS/IN2P3, France  \\
        E-mail: \email{holler@llr.in2p3.fr}}
\author{A.M.W. Mitchell\\
Max-Planck-Institut f{\"u}r Kernphysik, Heidelberg, Germany\\
E-mail: \email{alison.mitchell@mpi-hd.mpg.de}
}
\author{G. P{\"u}hlhofer\\
 Institut f\"ur Astronomie und Astrophysik, Universit\"at
T\"ubingen, T\"ubingen, Germany\\
E-mail: \email{Gerd.Puehlhofer@astro.uni-tuebingen.de}
}
\author{G. Rowell\\
School of Chemistry \& Physics, University of Adelaide,
Adelaide, Australia\\
E-mail: \email{gavin.rowell@adelaide.edu.au}
}
\author{S. Wagner\\
Landessternwarte, Universit\"at Heidelberg, K\"onigstuhl, Heidelberg,
Germany\\
E-mail: \email{swagner@lsw.uni-heidelberg.de}
}
  
\author{E. Bissaldi\\
INFN - Sez. di Bari, Via E. Orabona 4, Bari, Italy\\
E-mail: \email{Elisabetta.Bissaldi@uibk.ac.at}
}
\author{P. O'Brien\\
 Department of Physics and Astronomy, University of
Leicester, Leicester, United
Kingdom\\
E-mail: \email{paul.obrien@leicester.ac.uk}
}
\author{P.H.T. Tam\\
Institute of Astronomy and Space Science, Sun Yat-Sen University, Guangzhou, China\\
E-mail: \email{tanbxuan@sysu.edu.cn}
}
\author{on behalf of the H.E.S.S. Collaboration}

\abstract{ Gamma-ray bursts (GRBs) are some of the most energetic and
  exotic events in the Universe, however their behaviour at the
  highest energies (>10 GeV) is largely unknown. Although the
  \textit{Fermi}-LAT space telescope has detected several GRBs in this energy
  range, it is limited by the relatively small collection area of the
  instrument. The H.E.S.S. experiment has now entered its second phase
  by adding a fifth telescope of 600 m$^2$ mirror area to the centre of
  the array. This new telescope increases the energy range of the
  array, allowing it to probe the sub-100 GeV range while maintaining
  the large collection area of ground based gamma-ray observatories,
  essential to probing short-term variability at these energies.

  We will present a description of the GRB observation scheme used by
  the H.E.S.S. experiment, summarising the behaviour and performance
  of the rapid GRB repointing system, the conditions under which
  potential GRB repointings are made and the data analysis scheme used
  for these observations.  }

\FullConference{The 34th International Cosmic Ray Conference,\\
		30 July- 6 August, 2015\\
		The Hague, The Netherlands}

\begin{document}
%\linenumbers

%Begin a section.
\section{Introduction}

GRBs are the most luminous, highly-relativistic light sources known
and may be generated during the collapse of a massive star (long GRBs) or
via a merger event (short GRBs). They emit light across the
electromagnetic spectrum, including at GeV energies
(e.g. \cite{FermiCat}) and hence provide key targets for very high
energy (VHE) gamma-ray
detectors, such as
H.E.S.S. (e.g. \cite{HESSGRB,MAGICGRB,VERITASGRB,HAWCGRB}). During the
acceleration and emission period, GRBs may also be sources of
ultra-high-energy cosmic rays (UHECRs; \cite{Halzen}), in addition to
gravitational waves and neutrinos. Understanding the properties of
GRBs therefore permit multiple science objectives to be addressed
simultaneously. These include: (a) determine the physics of the jet
including the outflow speed and emission site; (b) search for hadronic
spectral signatures to explore jet composition and test the UHECR
origin; (c) constrain the particle acceleration mechanism and
radiation process(es); (d) measure the gamma-gamma attenuation due to
the extra-galactic background light; and (e) test for Lorentz
invariance violation (e.g. \cite{LorIn}). These objectives require
high photon rates and time-resolved spectroscopy, preferably extending
into the GeV-TeV band.

The H.E.S.S. gamma-ray telescope array has recently been upgraded by
the addition of a fifth, 600\,m$^2$ telescope to the array centre
(CT\,5). This large telescope greatly reduces the energy threshold of
the array.  Together with the large effective collection area, the
upgrade has greatly increased the chance for GRB detection. The
synergy of H.E.S.S. phase II with other multi-messenger facilities
would allow us to study GRBs in unparalleled ways.

\section{The H.E.S.S. II Rapid Repointing System}

\begin{figure*}
    \centering
	\includegraphics[width=0.5\textwidth]{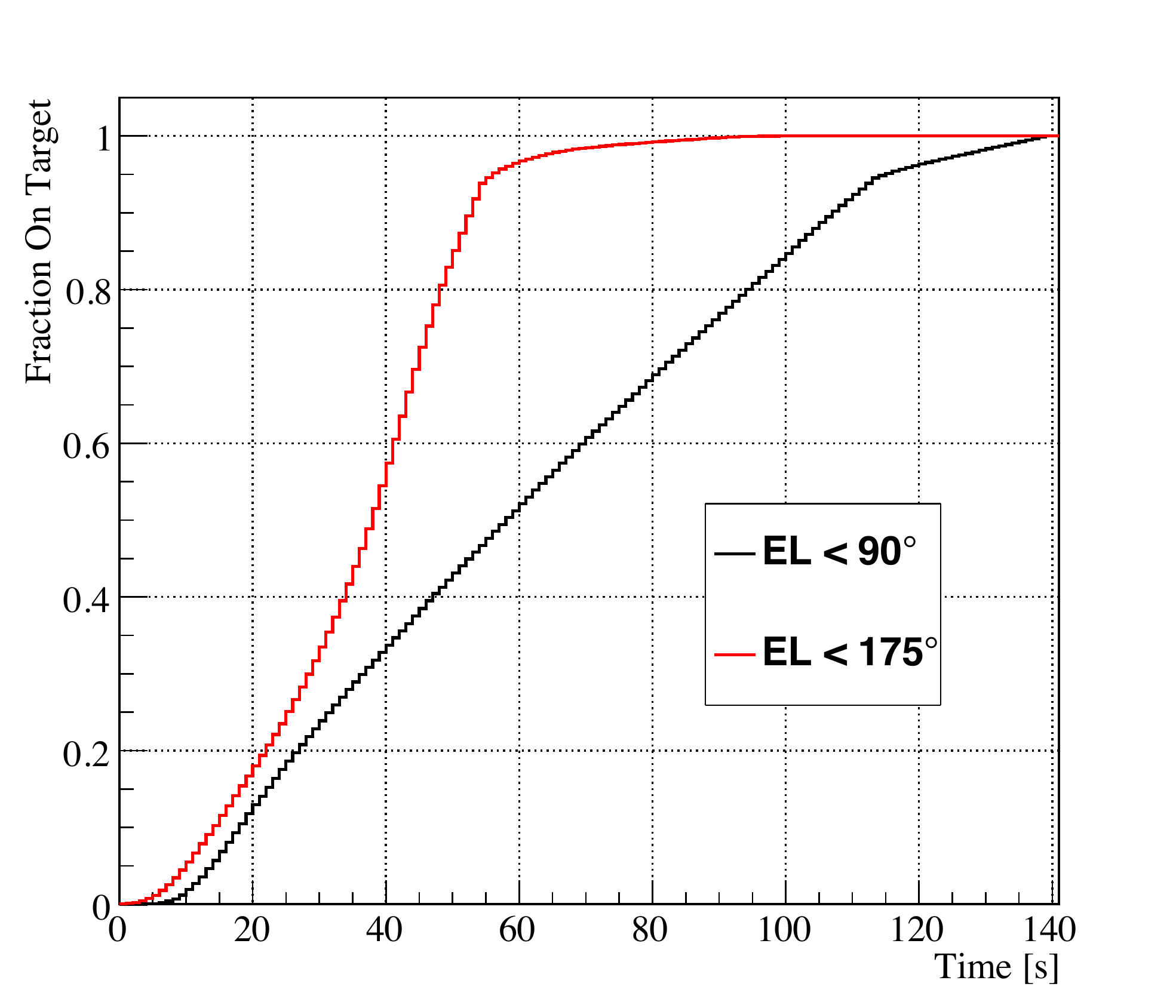}
	\caption{Fraction of times within which the CT5 telescope is
          able to arrive at a (random) target position on
          the sky versus the time after the issue of the repointing
          command. This fraction is shown for the systems with
          (elevation<175$^\circ$, red line) and without (elevation<90$^\circ$, black
          line) reverse-mode enabled. Figure reproduced from
          \cite{HESSIIdrive}. }
	\label{fig-drivesys}
\end{figure*}

As the gamma-ray luminosity of GRBs falls off very rapidly after burst
time, in order to maximise the chances of a GRB detection at very high
energies it is crucial to begin observations as soon as possible. In
order to minimise this delay two major improvements have been made for
the additional telescope (CT\,5) over the original 4 telescopes. Firstly the
telescope drive system of CT5 is significantly updated over that of
the original H.E.S.S. system (see \cite{HESSIIdrive} for details),
and is able to perform a full rotation of the telescope (180$^\circ$ in azimuth) in
$\sim$110 seconds. 
Additionally CT5 is permitted to point in \textit{reverse-mode}, allowing
the telescope to slew through zenith, resulting in significantly
faster repointing for some GRBs, where otherwise a large azimuthal
slew would be required (see Figure \ref{fig-drivesys}) .

In addition to this rapid slewing a fully automatic target of
opportunity (ToO) observation system has been implemented within the
H.E.S.S.  array, in this case receiving triggers from the GCN system.
The purpose of this system is to minimise the system repointing time
by reducing the number of hardware and software transitions required
to take place before observations can take place (see \cite{ArnimCHEP}
for a detailed explanation).

\begin{figure*}
    \centering
    \includegraphics[width=1.0\textwidth]{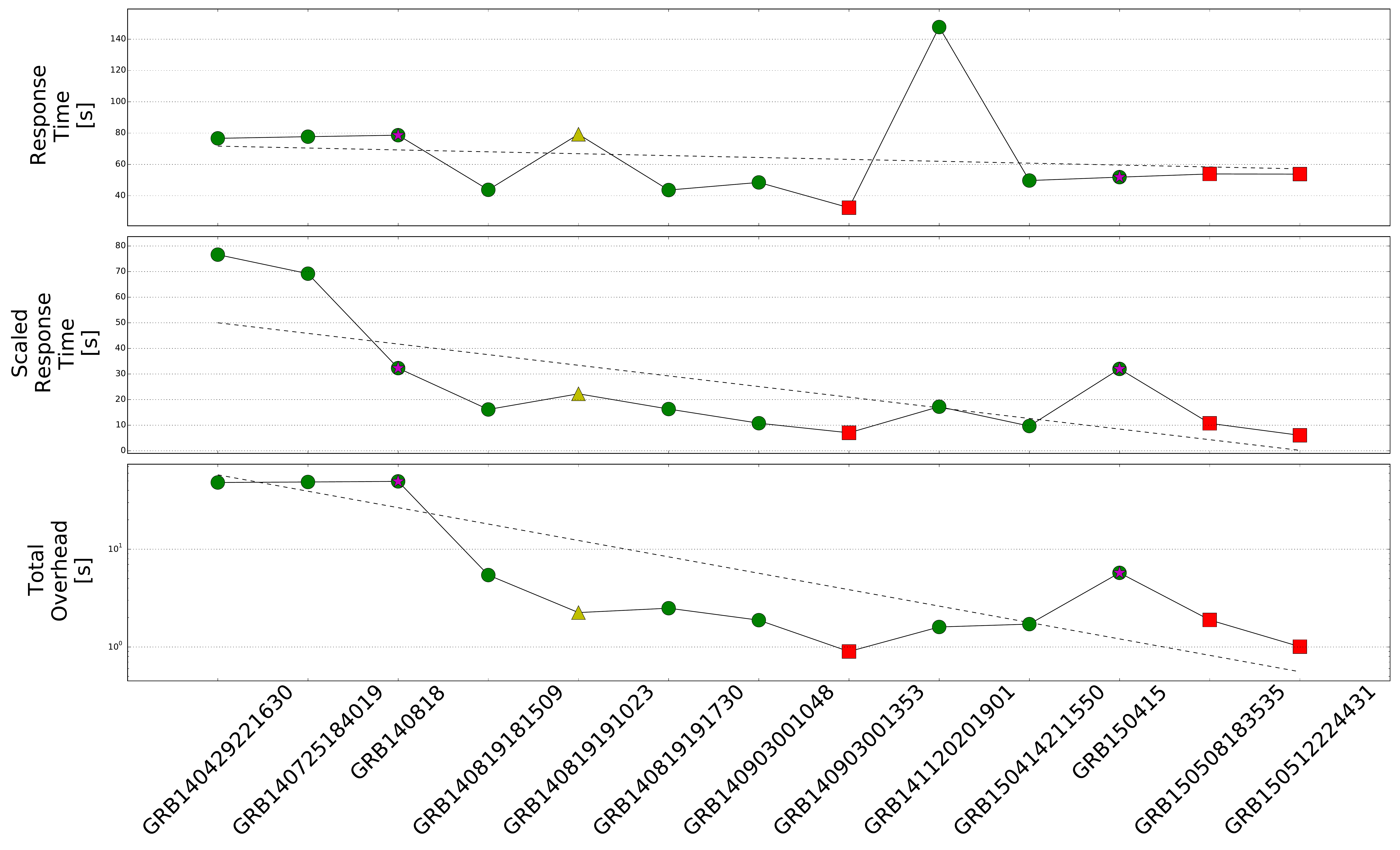}

	\caption{GRB observation total response time (top), response time
          scaled by the angular distance of the repointing (middle) and the
          data aquisition system time overhead (bottom) for a number of real and
          fake GRB observations. Figure adapted from
          \cite{ArnimCHEP}.}
	\label{fig-daqtimes}
\end{figure*}

The full ToO system is checked by the injection of fake GRB alerts at known
source positions on a monthly basis, in order to ensure the repointing
scheme is working reliably and the fastest possible repointing is
achieved. Figure \ref{fig-daqtimes} shows a log of the performance of
the ToO system on real and fake GRBs, demonstrating the fast
repositioning time and continual improvements to the system.

\section{GRB Observations with H.E.S.S. II}

\subsection{The GCN System}

The H.E.S.S. array receives the notification of potential
GRB repointings through the Gamma-ray Coordinates Network (GCN)
\cite{GCN}. The GCN system is designed to automatically distribute
details of GRB detections and localisations from a number of
instruments as rapidly as possible. Currently the H.E.S.S. array
receives GRB localisations from 2 instruments, the \textit{Fermi}-GBM and the
\textit{Swift}-BAT.

\subsubsection{\textit{Swift} Telescope}

The \textit{Swift} telescope is a multi-instrument telescope, launched in
2004, capable of rapid GRB triggers and follow-ups.
The Burst Alert Telescope (BAT) \cite{BAT} is a coded mask gamma-ray
detector sensitive in the 15-150 keV range. The BAT covers an angular range
of almost 1.4 steradians, and is able to provide localisations with a
positional accuracy at the arcminute level, with roughly 90 GRBs
detected per year. Typically bursts are followed up by the other
instruments on the \textit{Swift} telescope; the UV optical telescope (UVOT) and the
X-ray telescope (XRT), which provide spectra and light curves in these
wavelength bands. 

\subsubsection{\textit{Fermi} Gamma-ray Space Telescope}

The \textit{Fermi} Gamma-ray Burst Monitor (GBM) \cite{GBM} consists of a
series of scintillation detector placed around the \textit{Fermi} gamma-ray space
telescope providing almost all-sky coverage of gamma-ray transients in
the 8 keV to 40 MeV range, reporting to the GCN system around 200
GRB 
localisations per year. Although this instrument has excellent sky
coverage, unfortunately the accuracy of localisation is typically
somewhat limited, measuring the position of the GRB to an
accuracy of 2-5 degrees \cite{GBMacc}. Such a large error box can be
quite limiting when performing follow-up observations since it is is often of the order of the size of a typical IACT field of
view.
 
\subsection{Selection Criteria}

Selection of GRB follow-ups with H.E.S.S. are made on a combination of observability
based criteria and the likelihood of observable very high energy
emission from the object. GRB follow-up observations are split into
two categories, \textit{prompt} and \textit{afterglow}
observations. 

Prompt follow-ups are defined as those where the GRB occurs
within H.E.S.S. darktime and is immediately observable. In this case
the GRBs are subject only to observability criteria requiring that the
GRB position is observable for a minimum of 30 minutes above a zenith
angle of 60 degrees. Once an observation has begun, it will typically
continue for 2 hours, or until the source drops below the 60 degree
zenith threshold. If further, more accurate localisations are received
within the observation time the observation position is updated
automatically such that we are certain that the GRB is optimally
placed within the field of view. The expected rate of prompt
observation is around 4-6 per year of observations.

Conversely afterglow follow-ups are defined as GRB
localisations observable from the H.E.S.S. site, but not immediately
at burst time. Afterglow observations are subject to more stringent
observability cuts, requiring that the burst be observable above a zenith
angle of 45 degrees for 30 minutes or more. Additional cuts are placed
on the delay between the burst time and the time at which the GRB
becomes observable. In the most common case, where the redshift of the
GRB is unknown a maximum of 4 hours delay is allowed, this allowed
delay increases to 6 hours for sources with redshift below 1, 12 hours
for redshifts below 0.3, and 24 hours for redshifts below 0.1.
In the case of afterglows a GRB
expert is included in the decision making process, such that unusual
or especially interesting bursts can still be chosen
 The
expected rate of GRB follow-ups for these cuts is 8-12 observations
per year.

\section{Data Analysis Description}

\subsection{Data Blinding}

GRBs present an unusual observational situation for imaging
atmospheric Cherenkov telescopes (IACTs) in that they are transient
sources which in some cases may have a rather poor positional
accuracy. When searching for these sources we must be careful not to
introduce a large number of statistical trials while also being
certain that any detections are not a false positive. Therefore, in
order to ensure control of the trials factors the GRB observations are
subjected to a blinding procedure. After data taking is complete no
analysis of the data is made and the typical \textit{online} data
calibration and analysis is disabled, due to the large systematics
present in the automated procedure. Instead the data is transferred to
Europe blinded, where a \textit{Human in Loop} data calibration can
take place. However, efforts are ongoing to improve the low energy
stability of the on-site analysis, which may allow an accurate
automated data analysis in the future.

Once the data is calibrated each observation run is checked individually for known
camera hardware and data calibration issues, to ensure that no
spurious sources are generated in the final data analysis from these
issues.

\subsection{Data Analysis}

Once the data has been fully calibrated and checked it can then be
analysed using the H.E.S.S. II analysis pipeline. Currently events are
split into to two classes within the H.E.S.S. II system,
\textit{mono}-events which are seen only by CT\,5 and \textit{stereo}
events which are seen by at least 2 telescopes, these 2 classes can be
analysed separately or joined into a single \textit{combined} analysis
(full analysis scheme described in \cite{CrabICRC}). As the emission
spectrum of GRBs is expected to be quite steep, in this case the mono
event analysis (which has the lowest energy threshold) is expected to
be the most important. More details of the individual H.E.S.S. II
analysis schemes can be found in \cite{model,ImPACT,MonoReco}.

\section{Conclusions}

We have described the adaptations made to the H.E.S.S. II
observation scheme to allow for the fastest possible follow-ups of GRB
positions. Taking into account the delay of the GCN system, the
hardware and software overheads, and the repointing time of the array
we expect to arrive at the GRB target position within 2 minutes for
the majority of possible observation positions. Analysis of simulated
data shows that with the low energy optimised mono-event analysis we
expect to be sensitive down to around 30 GeV, greatly overlapping with the
observed range of energies already seen by Fermi-LAT \cite{LATGRBcat}, with a greatly
increased effective area. This rapid response, in conjunction with
good low energy performance should provide the H.E.S.S. array with a
strong chance of a future GRB detection at very high energies!

\vspace*{0.5cm}
\footnotesize{{\bf Acknowledgment:}{The support of the Namibian authorities and of the University of Namibia in facilitating the construction and operation of H.E.S.S. is gratefully acknowledged, as is the support by the German Ministry for Education and Research (BMBF), the Max Planck Society, the German Research Foundation (DFG), the French Ministry for Research, the CNRS-IN2P3 and the Astroparticle Interdisciplinary Programme of the CNRS, the U.K. Science and Technology Facilities Council (STFC), the IPNP of the Charles University, the Czech Science Foundation, the Polish Ministry of Science and Higher Education, the South African Department of Science and Technology and National Research Foundation, and by the University of Namibia. We appreciate the excellent work of the technical support staff in Berlin, Durham, Hamburg, Heidelberg, Palaiseau, Paris, Saclay, and in Namibia in the construction and operation of the equipment.}}

\end{document}